\documentclass{ws-procs975x65}

\begin{document}

\title{Charged black holes in Palatini $f(R)$ theories}

\author{G. J. Olmo$^1$ and D. Rubiera-Garcia$^2$}

\address{$^1$Departamento de F\'{i}sica Te\'{o}rica and IFIC,
Centro Mixto Universidad de Valencia - CSIC. Universidad de
Valencia, Burjassot-46100, Valencia, Spain \\
E-mail: gonzalo.olmo@csic.es}
\address{$^2$Departamento de F\'{i}sica, Universidad de Oviedo,
Avenida Calvo Sotelo 18, 33007, Oviedo, Asturias, Spain \\
E-mail: rubieradiego@gmail.com}

\begin{abstract}
In $f(R)$ extensions of General Relativity the Palatini approach provides ghost-free theories with second-order field equations and allows to obtain charged black hole solutions which depart from the standard Reissner-Nordstr\"om solution.
\end{abstract}


\bodymatter

\bigskip

$f(R)$ extensions of General Relativity (GR), defined by the action

\begin{equation}\label{eq:f(R)}
S=\frac{1}{2\kappa^2}\int d^4x\sqrt{-g} f(R)+S_M
\end{equation}
where $S_M$ is the matter action, admit two inequivalent formulations. In the standard metric formalism, where the connection is taken to be given by the Christoffel symbols of the metric, the modified dynamics is due to the existence of an extra (scalar) degree of freedom $\phi$, thus being equivalent to a scalar-tensor theory. In the Palatini approach, where metric and connection are taken to be independent, and the field equations obtained by variation of the action (\ref{eq:f(R)}) with respect to both of them, a scalar-tensor representation is also possible but the field $\phi$ carries no dynamics. Now the modified dynamics is due to a number of new terms that depend on the trace $T$ of the energy momentum tensor $T_{\mu\nu}$. In absence of matter, or for traceless energy-momentum matter sources, the dynamics boils down to that of GR with possibly a cosmological constant. This may be seen by deriving the field equations with no a priori relation between metric and connection:

\begin{eqnarray}
f_R R_{\mu\nu}-\frac{f}{2}g_{\mu\nu} &=& \kappa^2 T_{\mu\nu}\label{eq:met}\\
\nabla_{\beta}\left[\sqrt{-g}f_R g^{\mu\nu}\right]&=&0  \ ,
 \label{eq:con}
\end{eqnarray}
Taking the trace in (\ref{eq:met}) with $g^{\mu\nu}$ leads to the algebraic relation $Rf_R-2f=\kappa^2 T$, which generalizes the GR relation $R=-\kappa^2 T$ to $R=R(T)$. On the other hand, (\ref{eq:con}) can be seen as defining a new metric $h_{\mu\nu}$ such that $\nabla_{\beta}\left[\sqrt{-h} h^{\mu\nu}\right]=0$ and thus $\Gamma_{\mu\nu}^{\lambda}$ becomes the Levi-Civita connection of $h_{\mu\nu}$, i.e., $\Gamma^\lambda_{\mu \nu }=h^{\lambda \rho }(\partial_\mu h_{\rho \nu }+\partial_\nu h_{\rho \mu }-\partial_\rho h_{\mu \nu })/2$. Note that these two metrics are conformally related as $h_{\mu\nu}=f_R(T) g_{\mu\nu}$. In presence of matter with nonvanishing trace, the field equations (\ref{eq:met}) formulated in terms of $h_{\mu\nu}$ admit nontrivial solutions deviating from their GR counterparts; a conformal rescaling allows to put that solution in terms of the physical metric $g_{\mu\nu}$. Alternatively one may write the field equations directly in terms of $g_{\mu\nu}$

\begin{eqnarray}\label{eq:Gmn}
G_{\mu \nu }(g)&=&\frac{\kappa
^2}{f_R}T_{\mu \nu }-\frac{R f_R-f}{2f_R}g_{\mu \nu
}-\frac{3}{2(f_R)^2}[\partial_\mu f_R\partial_\nu
f_R-\frac{1}{2}g_{\mu \nu }(\partial f_R)^2]\nonumber \\
&+& \frac{1}{f_R}\left[\nabla_\mu \nabla_\nu f_R-g_{\mu \nu }\Box
f_R\right] \,
\end{eqnarray}
and solve them; this approach seems far too involved as compared to using $h_{\mu\nu}$ (though solutions to the analogous of (\ref{eq:Gmn}) in metric formalism have been obtained \cite{sol-fR}).

As an example we consider the case of nonlinear electrodynamics $\varphi(X,Y)$ \cite{f(R)}, defined as functions of the field invariants $X=-\frac{1}{2}F_{\mu\nu}F^{\mu\nu}$ and $Y=-\frac{1}{2}F_{\mu\nu}F^{*\mu\nu}$, where $F_{\mu\nu}=\partial_{\mu}A_{\nu}-\partial_{\nu}A_{\mu}$ and $F^{*\mu\nu}=\frac{1}{2} \epsilon^{\mu\nu\alpha\beta}F_{\alpha\beta}$. For electrostatic configurations $F^{tr}(r) \neq 0$ (for which $Y=0$) the trace $T=\frac{1}{2\pi} (\varphi-X\varphi_X)$ is nonvanishing (the only exception being Maxwell Lagrangian, $\varphi(X)=X$). The matter field equations read

\begin{equation} \label{eq:mat}
\varphi_X^2 X=q^2/r^4
\end{equation}
and form a compatible set with the $f(R)$ equations (\ref{eq:con}). In terms of $h_{\mu\nu}$ they read

\begin{equation} \label{eq:fieldequations}
{R_\mu}^\nu(h)=f_R^{-2}\Big(\kappa^2 {T_\mu}^\nu+\frac{f}{2}{\delta_\mu}^\nu \Big) \ ,
\end{equation}
For a spherically symmetric line element $ds^2=-g_{tt}dt^2+g_{rr}dr^2+r^2d\Omega^2=f_R^{-1}d\tilde{s}^2= f_R^{-1}(-A(\tilde{r})e^{\psi(\tilde{r})}dt^2+A(\tilde{r})^{-1}d\tilde{r}^2+\tilde{r}^2d\Omega^2)$ ($ds^2$ and $d\tilde{s}^2$ corresponding to $g_{\mu\nu}$ and $h_{\mu\nu}$, respectively) a explicit computation of Eqs.(\ref{eq:fieldequations}) with the ansatz for a mass function $A(\tilde{r})=1-2M(\tilde{r})/\tilde{r}$, and putting the result in terms of $g_{\mu\nu}$ leads to

\begin{equation}\label{eq:Mr}
\psi=0 \hspace{0.1cm}; \hspace{0.1cm} M_r=\left(4f_R^{3/2}\right)^{-1}\left(f+\frac{\kappa^2}{4\pi}\varphi\right)r^2\left(f_R+\frac{r}{2}f_{R,r}\right) \,
\end{equation}

It should be noted that higher-order curvature corrections ($R^2$, $R_{\mu\nu}R^{\mu\nu}$ and so on) arise in the quantization of fields in curved spacetimes \cite{quant}. Moreover, such terms also arise in several approaches to quantum gravity \cite{string} and can be motivated when GR is interpreted as an effective theory \cite{Cembranos}. In this sense, the $f(R)$ lagrangian $f(R)=R \pm l_P^2 R^2$, where $l_P \equiv$Planck's length can be regarded as the simplest modification of GR in this context. For the matter sector, the Born-Infeld lagrangian \cite{BI}

\begin{equation}
\varphi(X)=2\beta^2\left(1-\sqrt{1-\beta^{-2}X- (4\beta^4)^{-1} Y^2}\right) \ .
\end{equation}
originally introduced to remove the divergence of the electron's classical self-energy, is deemed of interest. From (\ref{eq:mat}) we obtain $\frac{X}{\beta ^2}=(1+z^4)^{-1}$, where $r^4= q^2z^4/\beta^2$. Since the GR relation $R=-\kappa^2 T$ holds, so that $\tilde{f}=-\tilde{T}+\lambda\tilde{T}^2$, with $\tilde{T}=T/\beta^2$ and $\lambda \equiv \kappa^2 \beta^2 l_P^2$, the mass function (\ref{eq:Mr}) in terms of the variable $z$ can be expressed as [$\gamma \equiv \sqrt{q/\beta}$]

\begin{equation}
\hat{M}(z)\equiv M(z)/M_0=1+(\gamma^3\kappa^2\beta^2/M_0) G(z)
\end{equation}
where $M_0$ is the Schwarzschild mass and the function $G(z)= -\int_z^\infty dz' \tilde{M}_{z'}$ [$\tilde{M}\equiv M/(\gamma^3\kappa^2\beta^2)$] contains the information on the geometry. The horizons (solutions of $A(r)=0$) are obtained as [$l_\beta^2\equiv 1/(\kappa^2\beta^2)$, $r_q^2\equiv \kappa^2 q^2/(4\pi)$, and $r_S\equiv 2M_0$]

\begin{equation}\label{eq:horizon}
1+2(4\pi)^{3/4}\left(r_q/r_S\right)\sqrt{r_q/l_\beta}G(z)=(4\pi)^{1/4}\left(r_q/r_S\right)\sqrt{l_\beta/r_q}zf_R^{1/2} \ .
\end{equation}
which highlights the three ratios (scales) present in the problem, namely, charge-to-mass $r_q/r_S$, NED-to-charge $l_\beta/r_q$, and NED-to-Planck $\lambda=l_P^2/l_\beta^2$. While the external horizon of these black holes occurs for $z\gg1$ and thus almost coincides with the GR value, near the center deviances from their GR counterparts are found.

A numerical analysis of the horizons equation (\ref{eq:horizon}) in terms of $\lambda$ ($\lambda=0$ the GR limit), assuming a small charge-to-mass ratio ($r_q/r_S\ll 1$) reveals several relevant modifications. i) For $f(R)=R+l_P^2 R^2$ there may be up to two inner horizons, which can be degenerate. ii) For $f(R)=R-l_P^2 R^2$ black holes have a similar structure in terms of horizons as the Reissner-Nordstr\"om one, but the radial coordinate cannot be extended below a minimum value $r_+$. iii) A curvature singularity always arise: the Kretschmann scalar $R_{\alpha\beta\gamma\delta}R^{\alpha\beta\gamma\delta}$ diverges as $\sim r^{-4}$ for $f(R)=R+l_P^2 R^2$ and as $\sim (r-r_+)^{-2}$ for $f(R)=R-l_P^2 R^2$, thus softening the central curvature singularity of the Reissner-Nordstr\"om solution ($\sim r^{-8}$). Thus, even though singularities are not removed in this theory, it seems a promising approach to the issue of black holes in modified gravity in Palatini approach, raising interest in more general quadratic Lagrangians such as those including Ricci-squared corrections \cite{or12a}.

This work has been supported by the Spanish grant FIS2011-29813-C02-02 and the JAE-doc program of the Spanish Research Council (CSIC).

\bibliographystyle{ws-procs975x65}
\bibliography{ws-pro-sample}

\end{document}